# The Geometry of the Universe

Wun-Yi Shu (許文郁)

Institute of Statistics, National Tsing Hua University, Taiwan
*Corresponding Author: shu@stat.nthu.edu.tw

**Abstract** In the late 1990s, observations of type Ia supernovae led to the astounding discovery that the universe is expanding at an accelerating rate. The explanation of this anomalous acceleration has been one of the great problems in physics since that discovery. We propose cosmological models that can simply and elegantly explain the cosmic acceleration via the geometric structure of the spacetime continuum, without introducing a cosmological constant into the standard Einstein field equation, negating the necessity for the existence of dark energy. In this geometry, the three fundamental physical dimensions length, time, and mass are related in new kind of relativity. There are four conspicuous features of these models: 1) the speed of light and the gravitational "constant" are not constant, but vary with the evolution of the universe, 2) time has no beginning and no end; i.e., there is neither a big bang nor a big crunch singularity, 3) the spatial section of the universe is a 3-sphere, and 4) in the process of evolution, the universe experiences phases of both acceleration and deceleration. One of these models is selected and tested against current cosmological observations, and is found to fit the redshift-luminosity distance data quite well.

## 1. Introduction

In the late 1990s, observations of type Ia supernovae made by two groups, the Supernova Cosmology Project [23] and the High-z Supernova Search Team [29], indicated that the universe appears to be expanding at an accelerating rate. The explanation of this anomalous acceleration has been one of the great problems in physics since that discovery. Currently, there are three mainstream approaches to explaining the accelerating expansion of the universe: the introduction of a cosmological constant; the proposal of scalar fields; and the developments of modified gravity. The first two approaches are commonly referred to as dark energy models. In the spatially flat $\Lambda$CDM model, presently the best-fit model to the available cosmological data, dark energy accounts for nearly three-quarters of the total mass-energy of the universe [3]. These approaches all raise several theoretical difficulties, and understanding the anomalous cosmic acceleration has become one of the greatest challenges of theoretical physics. Chapter 14 of Ellis, Maartens and MacCallum [11] provides a brief overview on this issue.

In this paper we propose cosmological models that can simply and elegantly explain the accelerating universe via the geometric structure of the spacetime continuum, without introducing a cosmological constant into the standard Einstein field equation, negating the necessity for the existence of dark energy. In this geometry, the three fundamental physical dimensions length, time, and mass are related in new kind of relativity. There are four conspicuous features of these models:

- The speed of light and the gravitational "constant" are not constant, but vary with the evolution of the universe.
- Time has no beginning and no end; i.e., there is neither a big bang nor a big crunch singularity.
- The spatial section of the universe is a 3-sphere, ruling out the possibility of a flat or hyperboloid geometry.
- In the process of evolution, the universe experiences phases of both acceleration and deceleration.

One of these models is selected and tested against current cosmological observations, and is found to fit the redshift-luminosity distance data quite well.

The paper is organized as the following: In the next section, the cosmological models are developed, with the details of the calculations presented in the Appendix. In Section 3, the dynamical evolution of the universe is determined by solving the field equation under various conditions. In Section 4, a selected model is tested against the observations of the type Ia supernovae contained in the Supernova Cosmology Project Union 2.1 Compilation (http://supernova.lbl.gov/Union/). Finally, the results are discussed in Section 5.

Throughout this paper we follow the sign conventions of Wald [31]. In particular, we use metric signature $-+++$, define the Riemann and the Ricci tensors by equations (3.2.3) and (3.2.25) of Wald [31] respectively, and employ abstract index notation to denote tensors. Greek indices, running from 0 to 3, are used to denote components of tensors while Latin indices are used to denote tensors. Einstein's summation convention is assumed.

## 2. Cosmological Models

A cosmological model is defined by specifying: 1) the spacetime geometry determined by a metric $g_{ab}$, 2) the mass-energy distributions described in terms of a stress-energy-momentum tensor $T_{ab}$, and 3) the interaction of the geometry and the mass-energy, which is depicted through a field equation.

### 2.1. The Spacetime Metric

Under the assumption that on the large scale the universe is homogeneous and isotropic, and expressed in the synchronous time coordinate and co-moving spatial spherical/hyperbolic coordinates $(t, \psi, \theta, \varphi)$, the line element of the spacetime metric $g_{ab}$ takes the form [31]

$$ds^2 = -c^2 dt^2 + a^2(t) \begin{cases} d\psi^2 + \sin^2\psi \left(d\theta^2 + \sin^2\theta\, d\varphi^2\right) \\ d\psi^2 + \psi^2 \left(d\theta^2 + \sin^2\theta\, d\varphi^2\right) \\ d\psi^2 + \sinh^2\psi \left(d\theta^2 + \sin^2\theta\, d\varphi^2\right) \end{cases}, \quad (1)$$

where $c$ is the speed of light and the three options listed to the right of the left bracket correspond to the three possible spatial geometries: a 3-sphere, 3-dimensional flat space, and a 3-dimensional hyperboloid respectively. The metric of form (1) is called the Friedmann-Robertson-Walker (FRW) metric.

We view the speed of light as simply a conversion factor between time and space. Like gravitation, it is a Nature's manifestation of the geometric structure of the spacetime continuum. Since the universe is expanding, this structure should be changing too, so we speculate that the conversion factor somehow varies in accordance with the evolution of the universe, hence the speed of light varies with the cosmic time. Denoting the speed of light as a function of the cosmic time by $c(t)$, we modify the FRW metric as

$$ds^2 = -c^2(t) dt^2 + a^2(t) \begin{cases} d\psi^2 + \sin^2\psi \left(d\theta^2 + \sin^2\theta\, d\varphi^2\right) \\ d\psi^2 + \psi^2 \left(d\theta^2 + \sin^2\theta\, d\varphi^2\right) \\ d\psi^2 + \sinh^2\psi \left(d\theta^2 + \sin^2\theta\, d\varphi^2\right) \end{cases}. \quad (2)$$

### 2.2. The Stress-energy-momentum Tensor

The universe is assumed to contain both matter and radiation. The content of the universe is described in terms of a stress-energy-momentum tensor $T^{ab}$. We shall take $T^{ab}$ to have the general perfect fluid form

$$T^{ab} = \left(\rho + \frac{P}{c^2}\right) u^a u^b + P g^{ab},$$

where $u^a$, $\rho$ and $P$ are, respectively, a time-like vector field representing the 4-velocity, the proper average mass-energy density, and the pressure as measured in the instantaneous rest frame of the cosmological fluid.

### 2.3. The Field Equation

In a cosmology where the speed of light $c$ and the Newtonian gravitational constant $G$ are assumed constant, the interaction between the curvature of spacetime at any event and the mass-energy content at that event is depicted through Einstein's field equation

$$G_{ab} \equiv R_{ab} - (1/2) R\, g_{ab} = 8\pi (G/c^4) T_{ab}, \quad (3)$$

where $R_{ab}$ is the Ricci tensor and $R$ is the curvature scalar. In a cosmology with a varying $c$ and a varying $G$, one needs a new field equation for attaining consistency; this is discussed in detail in Barrow [4] and Ellis and Uzan [12].

In a geometrical theory of gravity like general relativity, mass is measured in units of length. Noting that $G/c^2$ is the conversion factor that translates a unit of mass into a unit of length, we assert that $c$ and $G$ vary in such a way that $G(t)/c^2(t)$ must be absolutely constant with respect to the cosmic time $t$. Most of all, the constancy of $G/c^2$ can make equation (3) and the time-variations of $c$ and $G$ consistent with the usual meaning of conservation of mass-energy, as it will be shown that, with constancy of $G/c^2$, the vanishing covariant divergence of the right-hand side of (3) retains expressing the conservation of mass-energy. We can make $G(t)/c^2(t)=1$ by choosing proper units of mass and length. Accordingly, we obtain that in a cosmology with a varying $c$ and a varying $G$, the field equation describing the interaction between the spacetime geometry and the mass-energy content is given as

$$G_{ab} \equiv R_{ab} - (1/2) R\, g_{ab} = 8\pi (1/c^2) T_{ab} \equiv 8\pi\, T^*_{ab}, \quad (4)$$

where $T^*_{ab} \equiv (1/c^2) T_{ab}$. Due to the Bianchi identity, the contravariant tensor $T^{*\,ab}$ satisfies the equation of motion

$$\nabla_a T^{*\,ab} = 0.$$

In particular,

$$\nabla_\mu T^{*\,\mu\, 0} = 0,$$

which, after some straightforward algebra (see Appendix, Section A1 and A2), yields

$$\dot{\rho}(t) + 3\left[\rho(t) + \frac{P(t)}{c^2(t)}\right]\frac{\dot{a}(t)}{a(t)} = 0 .$$

Thus, for the universe composed of pressure-free dust ($P=0$), we have

$$\frac{d}{dt}\left[\rho(t)a^3(t)\right] = 0 ,$$

or equivalently,

$$\rho(t)a^3(t) = \text{constant} ; \qquad (5)$$

while for the universe composed of dust and radiation ($P = \rho c^2 / 3$), we have

$$\rho(t)a^4(t) = \text{constant} . \qquad (6)$$

Equations (5) and (6) express the conservation of proper mass-energy.

## 3. Dynamics of the Universe

In this section we determine the dynamical behavior, which is characterized by the functions $a(t)$ and $c(t)$ in metric (2), of the universe as described by our cosmological models. To obtain predictions for the dynamical evolution, we substitute the components of metric (2) into those of the field equation (4) and solve for $a(t)$ and $c(t)$. Computing the components of $G_{ab}$ in terms of $a(t)$ and $c(t)$, and then plugging the expressions for them into those of the field equation (4), after some tedious but straightforward calculation (see Appendix, Section A1 and A3), we arrive at the evolution equations for a homogeneous and isotropic universe:

$$\left[\frac{\dot{a}(t)}{c(t)}\right]^2 = \frac{8\pi}{3}\rho(t)a^2(t) - k , \qquad (7)$$

and

$$\frac{\ddot{a}(t)}{a(t)} - \frac{\dot{a}(t)}{a(t)}\frac{\dot{c}(t)}{c(t)} = -\frac{4\pi}{3}\left[\rho(t) + 3\frac{P(t)}{c^2(t)}\right]c^2(t) , \qquad (8)$$

where $k = 1$ for the 3-sphere, $k = 0$ for flat space, and $k = -1$ for the hyperboloid. Using equation (5), or respectively equation (6), we rewrite equation (7), for the universe composed of dust only, as

$$\left[\frac{\dot{a}(t)}{c(t)}\right]^2 = \frac{2M}{a(t)} - k , \qquad (9)$$

where $M \equiv 4\pi\rho(t)a^3(t)/3$, which is constant by equation (5); while for the universe composed of dust and radiation, as

$$\left[\frac{\dot{a}(t)}{c(t)}\right]^2 = \frac{2M'}{a^2(t)} - k ,$$

where $M' \equiv 4\pi\rho(t)a^4(t)/3$, which is constant by equation (6).

There are two unknown functions, $c(t)$ and $a(t)$, to be determined. To solve equations (8) and (9) we need a further postulate on the relationship between $c(t)$ and $a(t)$. For this we argue as follows: Time has no absolute meaning. The concept of time arises from the observation that the distribution of mass-energy contained in the universe is dynamic. There is no time apart from dynamicity. In the procedure of any measurement, one needs a standard to refer to. No matter how time is parameterized, the intrinsic length, in the geometry of spacetime continuum, of a span of time should be measured by the change in the mass-energy distribution during this period. The derivative, $\dot{\rho}(t)$, of the cosmological proper density is the very quantity that manifests the dynamicity of a homogeneous universe. When the magnitude of an increment in time, $dt$, is to be measured in units of length, $\dot{\rho}(t)$ must be taken to be the standard to refer to. If the distribution was static, i.e., $\dot{\rho}(t) \equiv 0$, the concept of time would have no meaning; in other words, the cosmological proper density plays the role of ultimate clock in a homogeneous universe. Accordingly, when being measured in units of length, the magnitude of an increment in time, $dt$, is renormalized with $|\dot{\rho}(t)|$. Therefore the conversion between time and length can be expressed as

$$dt \mapsto \left[\kappa_0 / |\dot{\rho}(t)|\right] dt ,$$

where $\kappa_0$ is constant with respect to the cosmic time $t$. Since the speed of light $c(t)$ is viewed simply as a conversion factor between time and length in the geometry of spacetime continuum, we also have the conversion

$$dt \mapsto c(t)dt .$$

Comparing the right hand sides of these two conversions, we conclude that

$$c(t) \propto 1/|\dot{\rho}(t)| .$$

Equation (5) gives,

$$\dot{\rho}(t) \propto \frac{\dot{a}(t)}{a^4(t)} .$$

Accordingly, we speculate that

$$c(t) = \kappa \left[ \frac{a^4(t)}{|\dot{a}(t)|} \right], \tag{10}$$

where $\kappa$ is constant with respect to the cosmic time $t$. We are now ready to solve equations (8) and (9) for $a(t)$ and $c(t)$. Given equations (5) and (10), equation (8) is redundant, so equation (9) is all we need to arrive at a solution. We will solve equation (9) for the universe composed of pressure-free dust and with spatially 3-sphere geometry (k=1) in detail and discuss the other cases briefly. Substituting (10) into equation (9) yields

$$\kappa^{-2} \left[ \frac{\dot{a}(t)}{a^2(t)} \right]^4 = \frac{2M}{a(t)} - 1. \tag{11}$$

Simplifying and preparing (11) for integration results in

$$\kappa^{-1/2} \dot{a}(t) = \pm [2M - a(t)]^{1/4} a(t)^{7/4}$$

$$\Rightarrow \kappa^{-1/2} \frac{da}{dt} = \pm (2M - a)^{1/4} a^{7/4}$$

$$\Rightarrow \kappa^{1/2} \frac{dt}{da} = \pm (2M - a)^{-1/4} a^{-7/4}$$

$$\Rightarrow dt = \pm \kappa^{-1/2} (2M - a)^{-1/4} a^{-7/4} da.$$

Carrying out the integration leads to

$$t(a) = \pm \kappa^{-1/2} \int_{2M}^{a} (2M - x)^{-1/4} x^{-7/4} dx$$

$$= \pm \frac{1}{2\kappa^{1/2} M} \int_{1}^{a/2M} (1-u)^{-1/4} u^{-7/4} du$$

$$= \pm \frac{2}{3\kappa^{1/2} M} \left( \frac{2M}{a} - 1 \right)^{3/4}, \quad 0 < a \le 2M. \tag{12}$$

We have chosen the time origin ($t=0$) to be the moment when $a$ achieves its maximum value $2M$. Setting $\sigma = 2/3\kappa^{1/2}M$ and solving equation (12) for $a$ in terms of $t$, we finally arrive at

$$a(t) = \frac{2M}{1 + (t/\sigma)^{4/3}}, \quad -\infty < t < \infty. \tag{13}$$

In this model, $a(t)$ is the hyper-radius of the universe at cosmic time $t$. The radius will get smaller and smaller as $t$ approaches $\pm\infty$, however it can never reach zero, and therefore, time has no beginning and no end, and there is neither a big bang nor a big crunch singularity. Setting $\gamma(t)=a(t)/2M$, the universe is accelerating in the epoch when $\gamma(t)<7/8$ and is decelerating when $\gamma(t)>7/8$. The graph of $a(t)/2M$ versus $t/\sigma$ is displayed in **Fig. 1**.

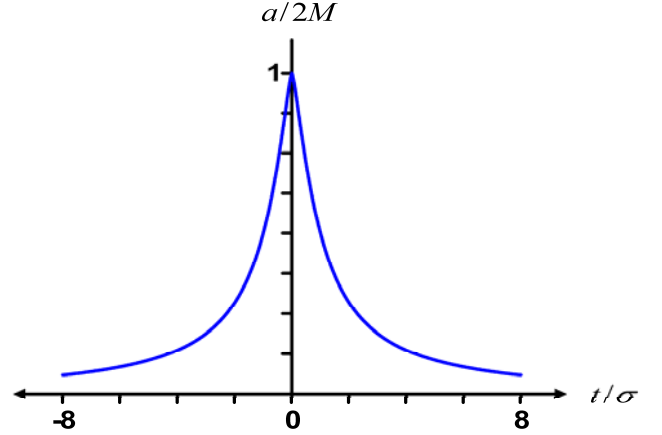

**Figure 1.** The evolution of the universe composed of pressure-free dust and with spatially 3-sphere geometry. The hyper-radius of the universe, $a(t)$, can never reach zero. The universe is accelerating in the epoch when $\gamma<7/8$ and is decelerating when $\gamma>7/8$.

From equations (10) and (13), the speed of light, as a function of the cosmic time $t$, can be calculated as

$$c(t) = \frac{(8M/3\sigma)}{\left[1 + (t/\sigma)^{4/3}\right]^2 (|t|/\sigma)^{1/3}}$$

$$= (4\kappa^{1/2} M^2) \gamma^2(t) \left[ \frac{1}{\gamma(t)} - 1 \right]^{-1/4}. \tag{14}$$

Since the speed of light $c$, wavelength $\lambda$, and frequency $\nu$ are related by $c=\lambda\nu$, a varying $c$ could be interpreted in different ways. We assume that a varying $c$ arises from a varying $\lambda$ with $\nu$ kept constant. We further assume that the relation between the energy $E$ of a photon and the wavelength $\lambda$ of its associated electromagnetic wave is given by equation $E(t)=\eta/\lambda(t)$, where $\eta$ is a constant that does not vary over cosmic time. Consequently, from relation $\lambda(t)=c(t)/\nu$, it follows that $E(t)=[\eta/c(t)]\nu\equiv h(t)\nu$. Therefore, the so called Planck's constant $h$ actually varies with the evolution of the universe.

Following the same procedure as above, the solutions for the other 5 cases are given as follows:

- For a universe composed of pressure-free dust and with spatially flat geometry,

$$t(a)/\sigma = (3/4) \int_{1}^{a/2M} u^{-7/4} du = 1 - \left( \frac{a}{2M} \right)^{-3/4},$$

$$0 < a < \infty.$$

$$a(t) = \frac{2M}{(1 - t/\sigma)^{4/3}}, \quad -\infty < t < \sigma.$$

We have chosen the time origin ($t=0$) to be the moment when $a$ reaches the value $2M$. In this case $a(t)$ will blow up at a finite future time $t=\sigma$. The graph of $a(t)/2M$ versus $t/\sigma$ is displayed in **Fig. 2**.

- For a universe composed of pressure-free dust and with spatially hyperboloid geometry,

$$t(a)/\sigma = (3/4) \int_1^{a/2M} (1+u)^{-1/4} u^{-7/4} du$$

$$= 2^{3/4} - \left(1 + \frac{2M}{a}\right)^{3/4}, \quad 0 < a < \infty. \quad (15)$$

Solving equation (15) for $a$ in terms of $t$ yields

$$a(t) = \frac{2M}{\left(2^{3/4} - t/\sigma\right)^{4/3} - 1},$$

$$-\infty < t < \sigma\left(2^{3/4} - 1\right).$$

In this case $a(0)=2M$ and $a(t)$ will blow up at a finite future time $t=\sigma(2^{3/4}-1)$. The graph of $a(t)/2M$ versus $t/\sigma$ is displayed in **Fig. 2**.

- For a universe composed of dust and radiation, and with spatially 3-sphere geometry,

$$t(a)/\sigma' = \pm \int_1^{a/\sqrt{2M'}} (1-u^2)^{-1/4} u^{-2} du,$$

$$0 < a \leq \sqrt{2M'},$$

or equivalently,

$$t(a)/\sigma' = \pm(1/2) \int_1^{a^2/2M'} (1-u)^{-1/4} u^{-3/2} du,$$

$$0 < a \leq \sqrt{2M'},$$

where

$$\sigma' \equiv \frac{1}{(\kappa')^{1/2}(2M')^{3/4}}, \text{ and } C(t) = \kappa' \left[\frac{a^5(t)}{|\dot{a}(t)|}\right].$$

We have chosen the time origin ($t=0$) to be the moment when $a$ achieves its maximum value $\sqrt{2M'}$. In this case $a(t) \sim |t|^{-1}$, as $t \to \pm\infty$.

- For a universe composed of dust and radiation, and with spatially flat geometry,

$$t(a)/\sigma' = \int_1^{a/\sqrt{2M'}} u^{-2} du = 1 - \frac{\sqrt{2M'}}{a},$$

$$a(t) = \frac{\sqrt{2M'}}{1 - t/\sigma'}.$$

In this case $a(0) = \sqrt{2M'}$ and $a(t)$ will blow up at a finite future time $t=\sigma'$.

- For a universe composed of dust and radiation, and with spatially hyperboloid geometry,

$$t(a)/\sigma' = \pm \int_1^{a/\sqrt{2M'}} (1+u^2)^{-1/4} u^{-2} du,$$

$$0 < a < \infty,$$

or equivalently,

$$t(a)/\sigma' = (1/2) \int_1^{a^2/2M'} (1+u)^{-1/4} u^{-3/2} du.$$

In this case $a(0) = \sqrt{2M'}$ and $a(t)$ will blow up at a finite future time before $t=\sigma'$.

From these results, we see that a spatially flat or spatially hyperboloid geometry is not feasible to describe our universe, since in either case $a(t)$ will blow up at a finite future time.

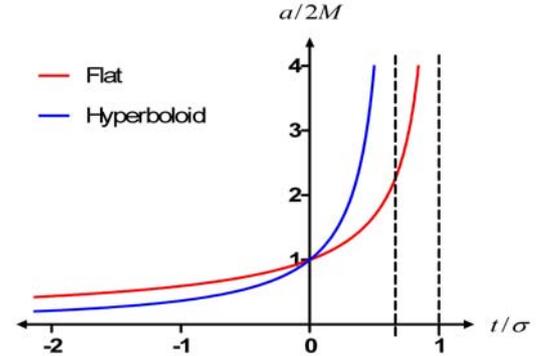

**Figure 2.** The dynamics of two versions of the universe composed of pressure-free dust: with spatially flat geometry, and with spatially hyperboloid geometry. In both universes, $a(t)$ can never reach zero and will blow up at a finite future time.

## 4. The Cosmological Redshift and Data Fitting

In this section we test the model for the universe composed of pressure-free dust and with spatially 3-sphere geometry against cosmological observations. Theoretical predictions of luminosity distance as a function of redshift will be compared with the observations of the type Ia supernovae contained in the Supernova Cosmology Project (SCP) Union 2.1 Compilation (http://supernova.lbl.gov/Union/).

Suppose that a photon of frequency (wavelength) $v_e$ ($\lambda_e$) is emitted at cosmic time $t_e$ by an isotropic observer $E$ with fixed spatial coordinates $(\psi_E, \theta_E, \varphi_E)$. Suppose this photon is observed at time $t_o$ by another isotropic observer $O$ at fixed co-moving coordinates. We may take $O$ to be at the origin of our spatial coordinate system. Let $v_o$ ($\lambda_o$) be the frequency (wavelength) measured by this second observer. The redshift factor, $z$, is given by

$$1 + z \equiv \frac{\lambda_o}{\lambda_e} = \frac{c(t_o)/v_o}{c(t_e)/v_e} = \frac{c(t_o)a(t_o)}{c(t_e)a(t_e)}.$$

Substituting the expression for $a(t)$ in (13) and that for $c(t)$ in (14) into the above equation yields

$$1 + z = \frac{a^3(t_o)}{a^3(t_e)} \left[\frac{2M}{a(t_o)} - 1\right]^{-1/4} \left[\frac{2M}{a(t_e)} - 1\right]^{1/4}. \quad (16)$$

By setting

$$\gamma(t) \equiv a(t)/2M, \; \gamma_e \equiv \gamma(t_e), \text{ and } \gamma_o \equiv \gamma(t_o),$$

equation (16) can be simplified as

$$1 + z = \gamma_o^{13/4} (1 - \gamma_o)^{-1/4} \gamma_e^{-13/4} (1 - \gamma_e)^{1/4}. \quad (17)$$

From metric (2) and the fact that $ds = d\theta = d\phi = 0$ along the photon path, we have

$$\psi_E = \int_{t_e}^{t_o} c(t)/a(t) \, dt.$$

Carrying out the integration (See Appendix, Section 4), leads to

$$\psi_E = 2\left[\tan^{-1}(1/\gamma_e - 1)^{1/2} - \tan^{-1}(1/\gamma_o - 1)^{1/2}\right]. \quad (18)$$

The proper distance between $E$ and $O$ at cosmic time $t_o$ is

$$d_P(z) = a(t_o)\psi_E,$$

and the luminosity distance between $E$ and $O$ at cosmic time $t_o$ is evaluated as

$$d_L(z) = a(t_o)(1 + z)\sin\psi_E$$
$$= 2M\gamma(t_o)(1 + z)\sin\psi_E. \quad (19)$$

In the SCP Union 2.1 Compilation, the luminosity distance is represented by the stretch-luminosity corrected effective B-band peak magnitude [23], $m_B^{effective}$, and the magnitude-redshift relation is given by

$$m_B^{effective}(z) = M_B + 5\log d_L(z) + 25,$$

where $M_B$ is a parameter believed to be constant for all supernovae of type Ia [24; 28; 30]. From equations (18) and (19), our model will predict the value $m_B^{effective}(z)$, as a function with two unknown parameters $\gamma_o$ and $\beta$, as follows:

$$m_B^{effective}(z)$$
$$= \beta + 5\log[\gamma_o(1+z)\sin\psi_E(\gamma_o, z)]$$
$$\equiv m_{\gamma_o,\beta}(z),$$

where $\beta \equiv M_B + 5\log 2M + 25$ and

$$\psi_E(\gamma_o, z)$$
$$= 2\left[\tan^{-1}\left(\frac{1}{\gamma_e(z)} - 1\right)^{1/2} - \tan^{-1}(1/\gamma_o - 1)^{1/2}\right]. \quad (20)$$

For a given $\gamma_o$, the quantity $\gamma_e(z)$ in (20), as a function of redshift factor $z$, is defined implicitly by equation (17). Plugging (20) into the expression for $m_{\gamma_0,\beta}(z)$ and after some tedious calculations (See Appendix, Section A4), yields:

$$m_{\gamma_0,\beta}(z)$$
$$\approx \xi_1 + 5\log\left[(1+z) - (1+z)^{11/13}\right]$$
$$\equiv m_{\xi_1}(z),$$

where

$$\xi_1 \equiv \beta + 5\log 2\gamma_o(1 - 2\gamma_o)\sqrt{\gamma_o(1 - \gamma_o)}.$$

The best-fit parameter is determined by minimizing the quantity

$$\sum_{i=1}^{n} d^2\left[(z_i, m_i), m_{\xi_1}(\bullet)\right]$$

over $\xi_1 \geq 0$, i.e.,

$$\xi_1^* = \arg\min_{\xi_1 \geq 0} \sum_{i=1}^{n} d^2\left[(z_i, m_i), m_{\xi_1}(\bullet)\right],$$

where $(z_i, m_i)$, $i = 1, 2, \ldots, n$, are the observations, and

$$d^2\left[(z_i, m_i), m_{\xi_1}(\bullet)\right] \equiv \min_{z \geq 0}\left[\left(\frac{z_i - z}{\sigma_{z_i}}\right)^2 + \left(\frac{m_i - m_{\xi_1}(z)}{\sigma_{m_i}}\right)^2\right].$$

**Fig. 3** shows the Hubble diagram of corrected effective rest-frame B magnitude as a function of redshift $z$ for the 714 supernovae contained in the test data set. The theoretical predictions, $m_{\xi_1}(z)$, of the model with $\xi_1$=28.341 fit the observations quite well.

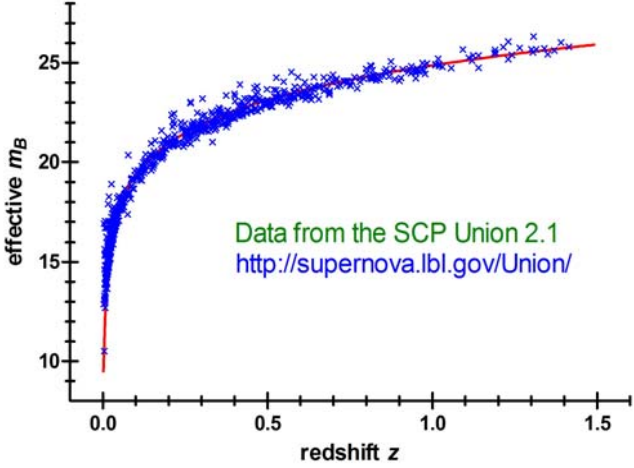

**Figure 3.** Hubble diagram for the 714 supernovae contained in the test data set. The solid red curve is the theoretical value of $m_{\xi 1}(z)$ as predicted by our model with parameter $\xi_1 = 28.341$

## 5. Discussions

In the Friedmann cosmology [13], a homogeneous and isotropic universe must have begun in a singular state. Hawking and Penrose [15] proved that singularities are generic features of cosmological solutions, provided only fairly general physical conditions. The prediction of singularities represents a breakdown of general relativity. Many people felt that the idea of singularities was repulsive and spoiled the beauty of Einstein's theory. There were therefore a number of attempts [5; 16; 17; 19] to avoid the conclusion that there had been a big bang, but were all abandoned eventually. Negating the existence of singularities restores beauty to Einstein's theory of general relativity.

Cosmological constant $\Lambda$ was introduced into the field equation of gravity by Einstein as a modification of his original theory to ensure a static universe. After Hubble's redshift observations [18] indicated that the universe is not static, the original motivation for the introduction of $\Lambda$ was lost. However, $\Lambda$ has been reintroduced on numerous occasions when it might be needed to reconcile theory and observations, in particular with the discovery of cosmic acceleration in the 1990s. With our models successfully explaining the accelerating universe without the introduction of $\Lambda$, the concept of cosmological constant shall be discarded again from the point of view of logical economy, as suggested by Einstein [10].

Beginning with Dirac [7] in 1937, some physicists have speculated that several so called physical constants may actually vary [6]. Theories for a varying speed of light (VSL) have been proposed independently by Petit [25; 26; 27] from 1988, Moffat [22] in 1993, and then Barrow [4] and Albrecht and Magueijo [1] in 1999 as an alternative way to cosmic inflation [2; 14; 20] of solving several cosmological puzzles such as the flatness and the horizon problems (for a detailed discussion of these problems, see Section 4.1 of Weinberg [32] and Section 9.7.1 of Ellis, Maartens and MacCallum [11]; for reviews of VSL, see Magueijo [21]). In the standard big bang cosmological models, the flatness problem arises from observation that the initial condition of the density of matter and energy in the universe is required to be fine-tuned to a very specific critical value for a flat universe. With our models asserting that the spatial section of the universe is a 3-sphere, the flatness problem disappears automatically. The horizon problem of the standard cosmology is a consequence of the existence of the big bang origin and the deceleration in the expansion of the universe. Without the big bang origin and with the universe being accelerating in the epoch when $\gamma(t) < 7/8$, our models may thus provide a solution to the horizon problem.

Essentially, this work is a novel theory about how the magnitudes of the three fundamental physical dimensions length, time, and mass are converted into each other, or equivalently, a novel theory about how the distribution of mass-energy and the geometry of spacetime interact. The theory resolves problems in cosmology, such as those of the big bang, dark energy, and flatness, in one fell stroke by postulating that

$$c(t) = \kappa \left[ \frac{a^4(t)}{|\dot{a}(t)|} \right] \quad \text{and}$$

$$\tau \equiv \frac{G(t)}{c^2(t)} = \text{constant}.$$

Since there are three fundamental physical dimensions, any cosmological model requires two constants to describe the relationship between them. Einstein took $c$ and $G$ as the two constants, whereas we assert that the two constants are $\kappa$, the factor relating to the conversion between time and length, and $\tau$, the conversion factor between mass and length. These two constants, $\kappa$ and $\tau$, together with $\eta$, the constant relating the energy of a photon and the wavelength of its associated electromagnetic wave, can be used to define the natural units of measurement for the three fundamental physical dimensions. Using dimensional analysis, we obtain:

- The natural unit of mass $\equiv \sqrt[4]{\dfrac{\eta}{\kappa \tau^3}}$

- The natural unit of time $\equiv \sqrt[4]{\dfrac{1}{\kappa \tau \eta}}$

- The natural unit of length $\equiv \sqrt[4]{\dfrac{\tau \eta}{\kappa}}$

Although, from expression (14), the speed of light becomes infinite at cosmic time $t=0$, it is of interest to note that

$$\int_{-\infty}^{\infty} c(t) dt = 2\pi M < \infty .$$

By comparing cosmological models, we refute the claim [8] that the time variation of a dimensional quantity such as the speed of light has no intrinsic physical significance. We illustrate our point as follows: In Friedmann's closed universe, which resulted from the constancy of the speed of light, the time span is a closed and bounded interval, from the big bang to the big crunch, while in ours the time span is an open interval, with neither beginning nor end. The two models can be discriminated by the topological structures of their time spans — the former is compact, whereas the latter is not. The mathematical definition of the concept of compactness, and the proof of the compactness of a closed and bounded interval can be found in Chapter XI of Dugundji [9]. Since compactness is a topological property, it's impossible to find a one-to-one and onto continuous correspondence between the time spans (for the proof, see Theorem 1.4, Dugundji [9], p. 224). This fact makes the two models intrinsically different.

## Acknowledgements

The author would like to thank Mr. Ching-Lung Huang (黃經龍) of the Molecular BioPhotonics Laboratory (MBPL) of the Department of Biomedical Engineering and Environmental Sciences, National Tsing Hua University, Taiwan, for designing computer programs for searching for the best parameter in data fitting and for providing the figures used in this paper.

## Appendix

**A1 Calculations for the Components of $\Gamma^c{}_{ab}$ and $R_{ab}$**

For the case of 3-sphere geometry, in the synchronous time coordinate and co-moving spatial spherical coordinates $(t, \psi, \theta, \varphi)$, the covariant components of metric $g_{ab}$, are

$$[g_{\mu\nu}] = \begin{bmatrix} -c^2(t) & 0 & 0 & 0 \\ 0 & a^2(t) & 0 & 0 \\ 0 & 0 & a^2(t)\sin^2\psi & 0 \\ 0 & 0 & 0 & a^2(t)\sin^2\psi\sin^2\theta \end{bmatrix},$$

and the contravariant components are

$$[g^{\mu\nu}] \equiv [g_{\mu\nu}]^{-1} = \begin{bmatrix} -c^{-2}(t) & 0 & 0 & 0 \\ 0 & a^{-2}(t) & 0 & 0 \\ 0 & 0 & a^{-2}(t)\sin^{-2}\psi & 0 \\ 0 & 0 & 0 & a^{-2}(t)\sin^{-2}\psi\sin^{-2}\theta \end{bmatrix}.$$

From equation (3.1.30) of Wald [31],

$$\Gamma^{\rho}_{\mu\nu} = \frac{1}{2}g^{\rho\sigma}\left(\partial_\mu g_{\nu\sigma} + \partial_\nu g_{\mu\sigma} - \partial_\sigma g_{\mu\nu}\right),$$

after some tedious but straightforward calculation, we find that the only non-vanishing components of the Christoffel symbol $\Gamma^c{}_{ab}$ are

$$\Gamma^0_{00} = \frac{\dot{c}(t)}{c(t)}, \quad \Gamma^0_{11} = \frac{a(t)\dot{a}(t)}{c^2(t)}, \quad \Gamma^0_{22} = \frac{a(t)\dot{a}(t)\sin^2\psi}{c^2(t)},$$

$$\Gamma^0_{33} = \frac{a(t)\dot{a}(t)\sin^2\psi\sin^2\theta}{c^2(t)},$$

$$\Gamma^1_{01} = \Gamma^1_{10} = \frac{\dot{a}(t)}{a(t)}, \quad \Gamma^1_{22} = -\sin\psi\cos\psi,$$

$$\Gamma^1_{33} = -\sin\psi\cos\psi\sin^2\theta,$$

$$\Gamma^2_{02} = \Gamma^2_{20} = \frac{\dot{a}(t)}{a(t)}, \quad \Gamma^2_{12} = \Gamma^2_{21} = \frac{\cos\psi}{\sin\psi},$$

$$\Gamma^2_{33} = -\sin\theta\cos\theta,$$

$$\Gamma^3_{03} = \Gamma^3_{30} = \frac{\dot{a}(t)}{a(t)}, \quad \Gamma^3_{13} = \Gamma^3_{31} = \frac{\cos\psi}{\sin\psi},$$

$$\Gamma^3_{23} = \Gamma^3_{32} = \frac{\cos\theta}{\sin\theta},$$

where the dots denote derivatives with respect to $t$. Substituting these expressions for the components of $\Gamma^c{}_{ab}$ into those for the components of the Ricci tensor $R_{ab}$,

$$R_{\mu\nu} = \partial_\beta \Gamma^\beta_{\mu\nu} - \partial_\mu \Gamma^\beta_{\beta\nu} + \Gamma^\alpha_{\mu\nu}\Gamma^\beta_{\alpha\beta} - \Gamma^\alpha_{\beta\nu}\Gamma^\beta_{\alpha\mu},$$

gives

$$R_{00} = 3\frac{\dot{a}(t)}{a(t)}\frac{\dot{c}(t)}{c(t)} - 3\frac{\ddot{a}(t)}{a(t)},$$

$$R_{11} = 2\left[\frac{\dot{a}(t)}{c(t)}\right]^2 + \frac{a(t)\ddot{a}(t)}{c^2(t)} - \frac{a(t)\dot{a}(t)\dot{c}(t)}{c^3(t)} + 2,$$

$$R_{22} = R_{11}\sin^2\psi, \quad R_{33} = R_{11}\sin^2\psi\sin^2\theta, \text{ and}$$

$$R_{\mu\nu} = 0, \text{ for } \mu \neq \nu.$$

From $R_\mu{}^\nu = R_{\mu\lambda}g^{\lambda\nu}$, it follows that

$$R_0{}^0 = 3\left[\frac{\ddot{a}(t)}{a(t)} - \frac{\dot{a}(t)\dot{c}(t)}{a(t)c(t)}\right]\frac{1}{c^2(t)}, \text{ and}$$

$$R_1{}^1 = R_2{}^2 = R_3{}^3 = 2\left[\frac{\dot{a}(t)}{a(t)c(t)}\right]^2 + \frac{\ddot{a}(t)}{a(t)c^2(t)} - \frac{\dot{a}(t)\dot{c}(t)}{a(t)c^3(t)} + \frac{2}{a^2(t)}.$$

Hence

$$R = R_\mu{}^\mu = \frac{6}{c^2(t)}\left[\frac{\ddot{a}(t)}{a(t)} + \frac{\dot{a}(t)^2}{a^2(t)} - \frac{\dot{a}(t)\dot{c}(t)}{a(t)c(t)} + \frac{c^2(t)}{a^2(t)}\right].$$

**A2 Equations of Motion**

Expressed in the synchronous time coordinate and co-moving spatial spherical/hyperbolic coordinates $(t, \psi, \theta, \varphi) \equiv [x^\mu]$, the 4-velocity of the fluid is simply $[u^\mu] = [\partial x^\mu / \partial t] = (1, 0, 0, 0)$. Hence the contravariant components of

$$T^{ab} = \left(\rho + \frac{P}{c^2}\right)u^a u^b + P g^{ab}$$

are

$$[T^{\mu\nu}] = \begin{bmatrix} \rho(t) & 0 & 0 & 0 \\ 0 & P(t)a^{-2}(t) & 0 & 0 \\ 0 & 0 & P(t)a^{-2}(t)\sin^{-2}\psi & 0 \\ 0 & 0 & 0 & P(t)a^{-2}(t)\sin^{-2}\psi\sin^{-2}\theta \end{bmatrix},$$

and the covariant components are

$$[T_{\mu\nu}] = \begin{bmatrix} \rho(t)c^4(t) & 0 & 0 & 0 \\ 0 & P(t)a^2(t) & 0 & 0 \\ 0 & 0 & P(t)a^2(t)\sin^2\psi & 0 \\ 0 & 0 & 0 & P(t)a^2(t)\sin^2\psi\sin^2\theta \end{bmatrix}.$$

Set $T^*_{ab} \equiv c^{-2}(t) T_{ab}$ and $T^{*ab} \equiv c^{-2}(t) T^{ab}$. From the contravariant version of the field equation

$$G^{ab} = 8\pi\, T^{*ab}$$

and the Bianchi identity, we see that $T^{*ab}$ satisfies the equation of motion

$$\nabla_a T^{*ab} = 0.$$

In particular,

$$\nabla_\mu T^{*\mu 0} \equiv T^{*\mu 0}{}_{;\mu} = 0.$$

From $\nabla_a T^{*bc} = \partial_a T^{*bc} + \Gamma^b_{ad} T^{*dc} + \Gamma^c_{ad} T^{*bd}$, we obtain

$$T^{*00}{}_{;0} = \partial_t T^{*00} + 2\Gamma^0_{00} T^{*00} = \frac{\dot{\rho}(t)}{c^2(t)},$$

$$T^{*10}{}_{;1} = \Gamma^1_{01} T^{*00} + \Gamma^0_{11} T^{*11}$$
$$= \left[\rho(t) + \frac{P(t)}{c^2(t)}\right]\frac{\dot{a}(t)}{a(t)}\frac{1}{c^2(t)},$$

$$T^{*20}{}_{;2} = \Gamma^2_{02} T^{*00} + \Gamma^0_{22} T^{*22}$$
$$= \left[\rho(t) + \frac{P(t)}{c^2(t)}\right]\frac{\dot{a}(t)}{a(t)}\frac{1}{c^2(t)},$$

and

$$T^{*30}{}_{;3} = \Gamma^3_{03} T^{*00} + \Gamma^0_{33} T^{*33}$$
$$= \left[\rho(t) + \frac{P(t)}{c^2(t)}\right]\frac{\dot{a}(t)}{a(t)}\frac{1}{c^2(t)}.$$

Therefore

$$T^{*\mu 0}{}_{;\mu} = \frac{\dot{\rho}(t)}{c^2(t)} + 3\left[\rho(t) + \frac{P(t)}{c^2(t)}\right]\frac{\dot{a}(t)}{a(t)}\frac{1}{c^2(t)},$$

which, together with $T^{*\mu 0}{}_{;\mu} = 0$, yields

$$\dot{\rho}(t) + 3\left[\rho(t) + \frac{P(t)}{c^2(t)}\right]\frac{\dot{a}(t)}{a(t)} = 0.$$

Thus, for the universe composed of pressure-free dust ($P=0$) we have

$$\dot{\rho}(t) + 3\rho(t)\frac{\dot{a}(t)}{a(t)} = 0$$

$$\Rightarrow \dot{\rho}(t)a^3(t) + 3\rho(t)a^2(t)\dot{a}(t) = 0$$

$$\Rightarrow \frac{d}{dt}\left[\rho(t)a^3(t)\right] = 0,$$

or equivalently,

$$\rho(t)a^3(t) = \text{constant}; \qquad (21)$$

while for the universe composed of dust and radiation ($P = \rho c^2 / 3$) we obtain

$$\dot{\rho}(t) + 4\rho(t)\frac{\dot{a}(t)}{a(t)} = 0$$

$$\Rightarrow \dot{\rho}(t)a^4(t) + 4\rho(t)a^3(t)\dot{a}(t) = 0$$

$$\Rightarrow \frac{d}{dt}\left[\rho(t)a^4(t)\right] = 0,$$

thus

$$\rho(t)a^4(t) = \text{constant}. \qquad (22)$$

## A3 The Evolution Equations for the Universe

Plugging the expression for $R$ and those for the components of $R_{ab}$, $g_{ab}$ and $T^*_{ab}$ into Einstein's field equation

$$R_{ab} - (1/2) R\, g_{ab} = 8\pi\, T^*_{ab},$$

since the three spatial field equations are equivalent due to the homogeneity and isotropy, we obtain just two equations:

$$\left[\frac{\dot{a}(t)}{a(t)}\right]^2 = \left[\frac{8\pi}{3}\rho(t) - \frac{1}{a^2(t)}\right] c^2(t), \quad (23)$$

and

$$\left[\frac{\dot{a}(t)}{a(t)}\right]^2 + 2\frac{\ddot{a}(t)}{a(t)} - 2\frac{\dot{a}(t)}{a(t)}\frac{\dot{c}(t)}{c(t)} = -\left(8\pi\frac{P(t)}{c^2(t)} + \frac{1}{a^2(t)}\right) c^2(t). \quad (24)$$

Using equation (23), we may rewrite equation (24) as

$$\frac{\ddot{a}(t)}{a(t)} - \frac{\dot{a}(t)}{a(t)}\frac{\dot{c}(t)}{c(t)} = -\frac{4\pi}{3}\left[\rho(t) + 3\frac{P(t)}{c^2(t)}\right] c^2(t).$$

Repeating the calculation for the cases of spatially flat and hyperboloid geometries, we obtain the general evolution equations for homogeneous, isotropic universe:

$$\left[\frac{\dot{a}(t)}{a(t)}\right]^2 = \left[\frac{8\pi}{3}\rho(t) - \frac{k}{a^2(t)}\right] c^2(t), \quad (25)$$

and

$$\frac{\ddot{a}(t)}{a(t)} - \frac{\dot{a}(t)}{a(t)}\frac{\dot{c}(t)}{c(t)} = -\frac{4\pi}{3}\left(\rho(t) + 3\frac{P(t)}{c^2(t)}\right) c^2(t),$$

where $k = 1$ for the 3-sphere, $k = 0$ for flat space, and $k = -1$ for the hyperboloid. Equation (25) can be rewritten as

$$\left[\frac{\dot{a}(t)}{c(t)}\right]^2 = \frac{8\pi}{3}\rho(t) a^2(t) - k.$$

Using equation (21) or, respectively, equation (22), we obtain, for the universe composed of dust only,

$$\left[\frac{\dot{a}(t)}{c(t)}\right]^2 = \frac{2M}{a(t)} - k,$$

where $M \equiv 4\pi\rho(t) a^3(t)/3$, which is constant by equation (21); while for the universe composed of dust and radiation,

$$\left[\frac{\dot{a}(t)}{c(t)}\right]^2 = \frac{2M'}{a^2(t)} - k,$$

where $M' \equiv 4\pi\rho(t) a^4(t)/3$, which is constant by equation (22).

## A4 The Magnitude-redshift Relation

The magnitude-redshift relation is given by equation

$$m_{\gamma_o,\beta}(z) = \beta + 5\log\left[\gamma_o(1+z) S(\psi_E)\right],$$

where

$$\psi_E = \int_{t_e}^{t_o} c(t)/a(t)\, dt,$$

$$c(t) = \kappa\left(a^4(t)\big/|\dot{a}(t)|\right), \text{ and}$$

$$1 + z \equiv \frac{\lambda_o}{\lambda_e} = \frac{c(t_o)/\nu_o}{c(t_e)/\nu_e} = \frac{c(t_o) a(t_o)}{c(t_e) a(t_e)}.$$

**(a)** For the case of 3-sphere geometry,

$$S(\psi_E) = \sin\psi_E, \quad a(t) = \frac{2M}{1 + (t/\sigma)^{4/3}},$$

where $\sigma = 2/3\kappa^{1/2} M$.

$$|\dot{a}(t)| = \left(\frac{8M}{3\sigma}\right)\left[\frac{a(t)}{2M}\right]^2 \left[\frac{2M}{a(t)} - 1\right]^{1/4},$$

$$1 + z = \frac{a^3(t_o)}{a^3(t_e)}\left[\frac{2M}{a(t_o)} - 1\right]^{-1/4}\left[\frac{2M}{a(t_e)} - 1\right]^{1/4}, \text{ and}$$

$$\frac{c(t)}{a(t)} = \kappa\left(\frac{a^3(t)}{|\dot{a}(t)|}\right) = \kappa^{1/2} a(t)\left[\frac{2M}{a(t)} - 1\right]^{-1/4}.$$

By letting

$$\gamma(t) \equiv a(t)/2M, \quad \gamma_e \equiv \gamma(t_e),$$
$$\text{and } \gamma_o \equiv \gamma(t_o),$$

it follows that

$$|\dot{a}(t)| = \left(\frac{8M}{3\sigma}\right)\gamma^2(t)\left[\frac{1}{\gamma(t)} - 1\right]^{1/4},$$

$$(1+z) = \gamma_o^{13/4}(1-\gamma_o)^{-1/4}\gamma_e^{-13/4}(1-\gamma_e)^{1/4}, \quad (26)$$

$$\frac{c(t)}{a(t)} = \left(2\kappa^{1/2}M\right)\gamma(t)\left[\frac{1}{\gamma(t)}-1\right]^{-1/4}, \text{ and}$$

$$dt = \frac{2M}{\dot{a}(t)}d\gamma(t)$$

$$= \left(\frac{3\sigma}{4}\right)\gamma^{-2}(t)\left[\frac{1}{\gamma(t)}-1\right]^{-1/4}d\gamma(t),$$

so we have

$$\psi_E = \int_{t_e}^{t_o}\gamma(t)^{-1/2}(1-\gamma(t))^{-1/2}d\gamma(t)$$

$$= \int_{\gamma_e}^{\gamma_o}u^{-1/2}(1-u)^{-1/2}du.$$

Carrying out the integration leads to

$$\psi_E = 2\left[\tan^{-1}(1/\gamma_e-1)^{1/2} - \tan^{-1}(1/\gamma_o-1)^{1/2}\right],$$

thus

$$m_{\gamma_o,\beta}(z) = \beta + 5\log\left[\gamma_o(1+z)\sin\psi_E(\gamma_o,z)\right],$$

where

$$\psi_E(\gamma_o,z) = 2\left[\tan^{-1}(1/\gamma_e(z)-1)^{1/2} - \tan^{-1}(1/\gamma_o-1)^{1/2}\right]. \quad (27)$$

For a given $\gamma_o$, the quantity $\gamma_e(z)$ in (27), as a function of redshift factor $z$, is defined implicitly by equation (26), which is equivalent to the following equations:

$$(1/\gamma_e)^{13/4}(1-\gamma_e)^{1/4}$$
$$= (1+z)(1/\gamma_o)^{13/4}(1-\gamma_o)^{1/4},$$

$$\left(\frac{1-\gamma_e}{1-\gamma_o}\right)^{1/4} = (1+z)\left(\frac{\gamma_e}{\gamma_o}\right)^{13/4}, \text{ and}$$

$$\left(\frac{\gamma_e}{\gamma_o}\right)^7(1+z)^2 = \sqrt{\frac{\gamma_e(1-\gamma_e)}{\gamma_o(1-\gamma_o)}}. \quad (28)$$

When $\gamma_o \approx 0$,

$$(1-\gamma_e)/(1-\gamma_o) \approx 1.$$

From (28) it implies that

$$\left(\frac{\gamma_e}{\gamma_o}\right)^7 \approx (1+z)^{-28/13}. \quad (29)$$

From expression (27), a straightforward calculation yields:

$$\sin\psi_E(\gamma_o,z)$$
$$= 2\left[(1-2\gamma_e)\sqrt{\gamma_o(1-\gamma_o)} - (1-2\gamma_o)\sqrt{\gamma_e(1-\gamma_e)}\right]$$

$$= 2(1-2\gamma_o)\sqrt{\gamma_o(1-\gamma_o)} \times$$
$$\left[\frac{1-2\gamma_e}{1-2\gamma_o} - \sqrt{\frac{\gamma_e(1-\gamma_e)}{\gamma_o(1-\gamma_o)}}\right],$$

which results in

$$m_{\gamma_0,\beta}(z)$$
$$= \beta + 5\log 2\gamma_o(1-2\gamma_o)\sqrt{\gamma_o(1-\gamma_o)} + 5\log(1+z)$$

$$+ 5\log\left[\frac{1-2\gamma_e}{1-2\gamma_o} - \sqrt{\frac{\gamma_e(1-\gamma_e)}{\gamma_o(1-\gamma_o)}}\right].$$

From (28) we obtain,

$$m_{\gamma_0,\beta}(z)$$
$$= \xi_1 + 5\log(1+z) + 5\log\left[\frac{1-2\gamma_e}{1-2\gamma_o} - \left(\frac{\gamma_e}{\gamma_o}\right)^7(1+z)^2\right],$$

where $\xi_1 \equiv \beta + 5\log 2\gamma_o(1-2\gamma_o)\sqrt{\gamma_o(1-\gamma_o)}$.

When $\gamma_o \approx 0$, by (29),

$$m_{\gamma_0,\beta}(z) \approx \xi_1 + 5\log(1+z) + 5\log\left[1-(1+z)^{-2/13}\right]$$
$$= \xi_1 + 5\log\left[(1+z)-(1+z)^{11/13}\right] \equiv m_{\xi_1}(z).$$

**(b)** For the case of flat geometry

$$S(\psi_E) = \psi_E, \quad a(t) = \frac{2M}{(1-t/\sigma)^{4/3}},$$

$$|\dot{a}(t)| = \left(\frac{8M}{3\sigma}\right)\left[\frac{a(t)}{2M}\right]^{7/4},$$

$$1+z = \left[\frac{a(t_o)}{a(t_e)}\right]^{13/4}, \text{ and}$$

$$\frac{c(t)}{a(t)} = \kappa\left[\frac{a^3(t)}{|\dot{a}(t)|}\right] = \kappa^{1/2}a(t)\left[\frac{a(t)}{2M}\right]^{1/4}.$$

By letting

$\gamma(t) \equiv a(t)/2M$, $\gamma_e \equiv \gamma(t_e)$,
and $\gamma_o \equiv \gamma(t_o)$,

it follows that

$$\left|\dot{a}(t)\right| = \left(\frac{8M}{3\sigma}\right)\gamma^{7/4}(t),$$

$$(1+z) = (\gamma_o/\gamma_e)^{13/4}, \qquad (30)$$

$$\frac{c(t)}{a(t)} = \left(2\kappa^{1/2}M\right)\gamma^{5/4}(t), \text{ and}$$

$$dt = \frac{2M}{\dot{a}(t)}d\gamma(t) = \left(\frac{3\sigma}{4}\right)\gamma^{-7/4}(t)d\gamma(t),$$

so we have

$$\psi_E = \int_{t_e}^{t_o} \gamma(t)^{-1/2} d\gamma(t)$$
$$= \int_{\gamma_e}^{\gamma_o} u^{-1/2} du = 2\left(\sqrt{\gamma_o} - \sqrt{\gamma_e}\right),$$

which results in

$$m_{\gamma_o,\beta}(z) = \beta + 5\log\left[2\gamma_o(1+z)\left(\sqrt{\gamma_o} - \sqrt{\gamma_e}\right)\right].$$

From (30) we obtain

$$m_{\gamma_o,\beta}(z)$$
$$= \beta + 5\log\left[2\gamma_o^{3/2}\right] +$$
$$+ 5\log\left[(1+z) - (1+z)^{11/13}\right]$$
$$= \xi_0 + 5\log\left[(1+z) - (1+z)^{11/13}\right] \equiv m_{\xi_0}(z),$$

where $\xi_0 \equiv \beta + 5\log\left[2\gamma_o^{3/2}\right]$.

**(c)** For the case of hyperboloid geometry

$$S(\psi_E) = \sinh\psi_E, \quad a(t) = \frac{2M}{\left(2^{3/4} - t/\sigma\right)^{4/3} - 1},$$

$$\left|\dot{a}(t)\right| = \left(\frac{8M}{3\sigma}\right)\left[\frac{a(t)}{2M}\right]^2 \left[\frac{2M}{a(t)}+1\right]^{1/4},$$

$$1+z = \frac{a^3(t_o)}{a^3(t_e)}\left[\frac{2M}{a(t_o)}+1\right]^{-1/4}\left[\frac{2M}{a(t_e)}+1\right]^{1/4}, \text{ and}$$

$$\frac{c(t)}{a(t)} = \kappa\left[\frac{a^3(t)}{|\dot{a}(t)|}\right] = \kappa^{1/2}a(t)\left[\frac{2M}{a(t)}+1\right]^{-1/4}.$$

By letting

$\gamma(t) \equiv a(t)/2M$, $\gamma_e \equiv \gamma(t_e)$,
and $\gamma_o \equiv \gamma(t_o)$,

it follows that

$$\left|\dot{a}(t)\right| = \left(\frac{8M}{3\sigma}\right)\gamma^2(t)\left[\frac{1}{\gamma(t)}+1\right]^{1/4},$$

$$(1+z) = \frac{\gamma_o^{13/4}(1+\gamma_o)^{-1/4}}{\gamma_e^{13/4}(1+\gamma_e)^{-1/4}}, \qquad (31)$$

$$\frac{c(t)}{a(t)} = \left(2\kappa^{1/2}M\right)\gamma(t)\left[\frac{1}{\gamma(t)}+1\right]^{-1/4}, \text{ and}$$

$$dt = \frac{2M}{\dot{a}(t)}d\gamma(t) = \left(\frac{3\sigma}{4}\right)\gamma^{-2}(t)\left[\frac{1}{\gamma(t)}+1\right]^{-1/4}d\gamma(t),$$

so we have

$$\psi_E = \int_{t_e}^{t_o} \gamma(t)^{-1/2}\left(1+\gamma(t)\right)^{-1/2} d\gamma(t)$$
$$= \int_{\gamma_e}^{\gamma_o} u^{-1/2}(1+u)^{-1/2} du.$$

Carrying out the integration leads to

$$\psi_E = 2\ln\left|\frac{\sqrt{1+\gamma_o} + \sqrt{\gamma_o}}{\sqrt{1+\gamma_e} + \sqrt{\gamma_e}}\right|,$$

thus

$$m_{\gamma_o,\beta}(z) = \beta + 5\log\left[\gamma_o(1+z)\sinh\psi_E(\gamma_o,z)\right], \text{ where}$$

$$\psi_E(\gamma_o,z) = 2\ln\left|\frac{\sqrt{1+\gamma_o} + \sqrt{\gamma_o}}{\sqrt{1+\gamma_e(z)} + \sqrt{\gamma_e(z)}}\right|. \quad (32)$$

For a given $\gamma_o$, the quantity $\gamma_e(z)$ in (32), as a function of redshift factor $z$, is defined implicitly by equation (31), which is equivalent to the following equations:

$$(1/\gamma_e)^{13/4}(1+\gamma_e)^{1/4} = (1+z)(1/\gamma_o)^{13/4}(1+\gamma_o)^{1/4},$$

$$\left(\frac{1+\gamma_e}{1+\gamma_o}\right)^{1/4} = (1+z)\left(\frac{\gamma_e}{\gamma_o}\right)^{13/4}, \text{ and}$$

$$\left(\frac{\gamma_e}{\gamma_o}\right)^7 (1+z)^2 = \sqrt{\frac{\gamma_e(1+\gamma_e)}{\gamma_o(1+\gamma_o)}}. \quad (33)$$

When $\gamma_o \approx 0$, $(1+\gamma_e)/(1+\gamma_o) \approx 1$, from (33) it implies that

$$\left(\frac{\gamma_e}{\gamma_o}\right)^7 \approx (1+z)^{-28/13}. \quad (34)$$

From expression (32), a straightforward calculation yields:

$$\sinh\psi_E(\gamma_o,z) = \left(\frac{1}{2}\right)\left[\left(\frac{\sqrt{1+\gamma_o}+\sqrt{\gamma_o}}{\sqrt{1+\gamma_e}+\sqrt{\gamma_e}}\right)^2 - \left(\frac{\sqrt{1+\gamma_e}+\sqrt{\gamma_e}}{\sqrt{1+\gamma_o}+\sqrt{\gamma_o}}\right)^2\right].$$

When $\gamma_o \approx 0$,

$$\left(\frac{\sqrt{1+\gamma_o}+\sqrt{\gamma_o}}{\sqrt{1+\gamma_e}+\sqrt{\gamma_e}}\right)^2 \approx \left[1+2\sqrt{\gamma_o(1+\gamma_o)}\right] \times \left[1-2\sqrt{\gamma_e(1+\gamma_e)}\right],$$

$$\left(\frac{\sqrt{1+\gamma_e}+\sqrt{\gamma_e}}{\sqrt{1+\gamma_o}+\sqrt{\gamma_o}}\right)^2 \approx \left[1+2\sqrt{\gamma_e(1+\gamma_e)}\right] \times \left[1-2\sqrt{\gamma_o(1+\gamma_o)}\right].$$

Therefore

$$\sinh\psi_E(\gamma_o,z) \approx 2\left[\sqrt{\gamma_o(1+\gamma_o)} - \sqrt{\gamma_e(1+\gamma_e)}\right]$$
$$= 2\sqrt{\gamma_o(1+\gamma_o)}\left[1 - \sqrt{\frac{\gamma_e(1+\gamma_e)}{\gamma_o(1+\gamma_o)}}\right].$$

By (33) and (34), we have

$$\sinh\psi_E(\gamma_o,z) \approx 2\sqrt{\gamma_o(1+\gamma_o)}\left[1-(1+z)^{-2/13}\right],$$

which results in

$$m_{\gamma_0,\beta}(z)$$
$$\approx \beta + 5\log\left[2\gamma_0\sqrt{\gamma_o(1+\gamma_o)}\right] + 5\log\left[(1+z)-(1+z)^{11/13}\right]$$
$$= \xi_{-1} + 5\log\left[(1+z)-(1+z)^{11/13}\right] \equiv m_{\xi_{-1}}(z),$$

where $\xi_{-1} \equiv \beta + 5\log\left[2\gamma_0\sqrt{\gamma_o(1+\gamma_o)}\right]$.

**Summary**: The magnitude-redshift relation is given by

$$m_{\xi_k}(z) = \xi_k + 5\log\left[(1+z)-(1+z)^{11/13}\right],$$

$$\xi_k \equiv \begin{cases} \beta + 5\log\left[2\gamma_o(1-2\gamma_o)\sqrt{\gamma_o(1-\gamma_o)}\right], & \text{if } k=1 \\ \beta + 5\log\left(2\gamma_o^{3/2}\right), & \text{if } k=0 \\ \beta + 5\log\left[2\gamma_0\sqrt{\gamma_o(1+\gamma_o)}\right], & \text{if } k=-1 \end{cases},$$

where $k=1$ for the 3-sphere, $k=0$ for the flat, and $k=-1$ for the hyperboloid geometry respectively, and $\beta \equiv M_B + 5\log 2M + 25$.